# Electronic viscosity in a multiple quantum well system

Partha Goswami[1], Ajay Pratap Singh Gahlot[1] and Avinashi Kapoor[2]

[1] D.B.College, University of Delhi, Kalkaji, New Delhi-110019, India
[2] Department of Electronic Science, University of Delhi south campus, New Delhi-110023, India

**Absract** We calculate the electronic viscosity ($\eta$) for a multiple quantum well structure in the presence of disorder potential(V ~ 4 meV), the electron-electron repulsion($U_0$~ 5-17 meV) and a strong magnetic field (B $\geq$ 16.5 T) in the direction in which the electrons are trapped. The quantity $\eta$ is different from the dissipation-less Hall viscosity ($\eta_H$) which cannot take non-zero value in a time reversal invariant system. The Fermi energy density of states for the system has been calculated in the t-matrix approximation assuming low concentration of impurities. Our approach involves calculation of the density of viscosity $\eta(\mathbf{k})$, for temperature close to 0 K, on the Brillouin zone(BZ) followed by the numerical evaluation of the integral of $\eta(\mathbf{k})$ over the BZ. We show that (i) $\eta$ is nearly proportional to B for given (V,$U_0$),and (ii) dissipation-less state($\eta\to 0^+$), analogous to superfluidity, is possible for a critical value of $U_0$ when (V, B) are given. We also calculate the entropy per particle (S) and show that the results comply with the KSS bound $\eta/S \geq \hbar/4\pi k_B$ reported by Kovtun et al. [P.K. Kovtun, D.T. Son, and A.O.Starinets, Phys. Rev. Lett. 94, 111601 (2005); Taro Kimura, arXiv: 1004.2688(Unpublished)].

**1 Introduction** In this paper we present an account of the electronic viscosity calculation in the (momentum)spatio-(imaginary) temporal formulation of the multiple quantum well(MQW) problem involving the Landau level(LL) split sub-band energies $\varepsilon_\alpha$ ($\alpha$ = 0,1), the disorder potential(V ~ 4 meV) and the electron-electron repulsion($U_0$ = 5-10 meV). It must be emphasized that the picture here corresponds to a fixed areal density ($N_A$)of the carriers, and a changing magnetic field (B).The quantity $N_A$ has been calculated within the density functional theory(DFT)[1] based iterative computational scheme assuming a suitable parabolic confinement in the z-direction of the form V(z) = [ $-V_0$ + (1/2) $m_z$ $\omega_0^2$ $z^2$ ] for $-d/2 < z < d/2$ (and zero elsewhere) where the bulk band-gap mismatch is $V_0$ between the well and the barrier materials and the well-layer width is d. We find that $\omega_0 = (4V_0/m_z d^2)^{1/2}$ where $m_z$ is the tranverse electronic mass. Upon taking $V_0$ = 0.35 eV, d = 10 nm, and barrier width b = 20 nm, we find $\varepsilon_0 = -277.2$ meV, $\varepsilon_1 = -131.6$ meV and $N_A = 4\times10^{15}$ m$^{-2}$. Now the maximum number of particles with spin s = ½per Landau level (LL) is given by D = 2(Be/h) A, where A is the well-layer area. Since the filling factor($\nu$) = ($N_A$ h/eB), one finds D = (20/$\nu$) with the number of current carrying states in the single well-layer area A (A = 50 $\times$ 50 nm$^2$) to be ($N_A$A) = 10. The number of LLs considered with each sub-band is not more than two (indexed by N=0,1). Obviously enough, for the entire MQW system with j well-layers the total number of available current carrying states is 10j. Furthermore, for $\nu$ =1, from $\nu$ = ($N_A$ h/eB), one finds B = $B_0$ = 16.5 T and, with V= 4 meV, $U_0$ = 5 meV and T = 3 K, the chemical potential of the fermion number $\mu = -262.35$ meV which is closer to the sub-band $\varepsilon_0$. As $U_0$ is increased, $\mu$ exhibits a marginal shift (e.g, for $U_0$ = 10 meV, $\mu = -270.46$ meV). As $\nu$ decreases, due to increase in B, there is again a marginal shift in $\mu$. For example for $\nu$ = 3/7 and 1/3,respectively, $\mu = -260.81$ meV and $-260.17$ meV when $U_0$ = 5 meV. It may be mentioned that Khrapai et al.[2] have reported the chemical potential jump for electrons at fractional filling factors. We have determined entropy per particle(S) and viscosity($\eta$) using the values of $\mu$ obtained. The universal relation[3,4] between S and $\eta$, viz. $\eta/S \geq \hbar/4\pi k_B$ (=6.08× $10^{-13}$ K-s), is complied with for fractional filling factors. The approximation made to treat the coulombs repulsion is reminiscent of the well-known Hubbard approximation[5]. We now clarify in brief how the chemical potential and viscosity have been determined and also make an effort to articulate the physical motivation behind this work.

In order to calculate the viscosity we need to know the single-particle Green's functions with the inclusion of the effect of elastic scattering by impurities. For this, we take into consideration a real space Hamiltonian $H_{rs}$- a tight binding description of the system including the intra-layer electron-electron interaction, the confinement potential V(z), etc.. We have assumed here the inter-layer electron-electron interaction to be zero which allowed us to separate the Hamiltonian $H_{rs}$ into two parts describing respectively the motion in the z-direction and the motion of the two-dimensional electron gas (2DEG) in the x-y plane. Obviously, the inter-layer Coulomb interaction couples these two subsystems in a non-trivial way. This is followed by the crystal electron approximation (dividing the single well layer area into $N_s = N \times M$ square unit cells each of side length 'a') and the application of the second quantization formalism. The expressions for the quasi-particle field operators (QPFO) at any point in a well-layer, thus obtained, would involve the coefficients which are the single-particle wave functions possessing the required property of Bloch waves and sum $\sum_{k,\alpha,N}$ in QPFO is over the complete set of single-particle quantum numbers $k(\equiv(\mathbf{k},\sigma)),\alpha,$ and N. The creation/ destruction operator amplitudes for the states specified by $(\mathbf{k},\alpha,N,\sigma)$ are, say, $a^{\dagger}_{k,\alpha,N,\sigma}$ / $a_{k,\alpha,N,\sigma}$. We worked out what would approximately be the measure of the overlap for intra/ inter sub-band electron wave functions, considering a minimal description for a particle with mass m and charge e on a plane in a transverse magnetic field B with the vector potential A assumed to be given in the symmetric gauge $\mathbf{A} = B(-y/2 \quad x/2 \quad 0)$ and calculating the ground state and the excited state eigenfunctions. For example, the numerical values of the intra sub-band electron wave function overlap with the first and the second neighbours, respectively, have been estimated to be t ~ 1 meV, and $t'$ ~ 0.1t. The steps outlined above lead us to a second-quantized Hamiltonian $H_{MQW}$ in momentum space (see Eqs.(1) and (2)) for the z-direction confined Bloch electrons of spin $\sigma$ in the well-layers parallel to x-y plane. Since variants of the calculations above could be found in the standard text-books on non-relativistic field theoretic methods for many-body problems [6,7], we refrain from giving the details. The single-particle Green's function $G_{\alpha,N;\alpha',N',\sigma}$ $(\mathbf{k},\tau) = -\langle T\{ a_{k,\alpha,N,\sigma}(\tau) a^{\dagger}_{k,\alpha',N',\sigma}(0)\}\rangle$ calculated on the basis of this Hamiltonian leads to excitation spectra E($\mathbf{k}$).

The impurity potential/disorder with finite range has drastic effects on the density of states at Fermi energy relevant for thermodynamic and transport properties. Although it might be thought that a perfect crystal would give the strongest quantum Hall effect (QHE), the effect actually relies on the presence of dirt in the samples. In practice, the QHE related phenomena are strongly affected by the presence of even weak disorder, so any practical discussion of these phenomena must be preceded by an analysis of the effects of disorder. For this reason we have included the effect of elastic scattering by impurities. We assume that impurities are alike, distributed randomly, and contribute a momentum dependent potential of the form $V(|\mathbf{q}|) = [|v_0|^2 \kappa^2 /\{|\mathbf{q}|^2 + \kappa^2\}]^{1/2}$, where $\kappa^{-1}$ characterizes the range of the impurity potential. The limit $\kappa \gg |\mathbf{q}|$, which corresponds to a point-like isotropic scattering potential characterizing the in-plane impurities, will only be considered here for simplicity. The elastic scattering by impurities involves the calculation[6] of self-energy $\Sigma(\mathbf{k},\omega_n)$ in terms of the momentum and the Matsubara frequencies $\omega_n = [(2n+1) \pi k_B T]$, n = 0, ± 1, ± 2,……. This alters the single-particle excitation spectra $E(\mathbf{k})$ in a fundamental way. We obtain renormalized Matsubara propagators $G^{(ren)}_{\alpha,N;\alpha',N',\sigma}(\mathbf{k},\omega_n)$ in the t-martix approximation[6]. The approximation is good enough for low impurity concentration. The propagators yield the density of states(DOS)$\rho(\mathbf{k},\omega)$ as a Lorentzian. The equation to determine $\mu$ is given by analog of the Luttinger sum-rule (for the situation when the magnetic field is kept fixed and the carrier density is changed) involving the thermal averages $\langle a^{\dagger}_{\mathbf{k},\alpha',N',\sigma} a_{\mathbf{k},\alpha,N,\sigma} \rangle$ which are obtainable from $G^{(ren)}_{\alpha,N;\alpha',N',\sigma}(\mathbf{k},\omega_n)$. The Fermi energy DOS is computed numerically from $\rho(\mathbf{k},\omega)$ putting $\omega = \mu$. In the same units as those of Kovtun et al.[3,4], the viscosity is defined in Eq. (17) below in terms of density of viscosity $\eta(\mathbf{k})$ involving $\rho(\mathbf{k},\omega = \mu)$. To evaluate the **k**-summation here, we divide the first Brillouin zone(BZ) into finite number of grids. We next determine the numerical values, corresponding to each of these grids, of the density $\eta(\mathbf{k})$ and sum these values. We have generated these values through the surface plot of $\eta(\mathbf{k})$ using 'Matlab' software. The plot resolution is throughout kept at 0.0281 which gives rise to 50176 grids.

In the present two sub-band scenario, the mean in-plane carrier spacing($a_0=(2\pi N_A)^{-1/2}$ )is 6.3078 nm and the magnetic length $l_B = \sqrt{(\hbar/eB)} \geq 6$ nm(for B ≤ 16.5 T). For much stronger magnetic field one would have $l_B < a_0$. Furthermore, the characteristic length - the confinement width of the 2D electron wave function $z_0 = (\hbar d/\sqrt{(4m_z V_0)})^{1/2} \sim 1.5$ nm. This is the effective width ($z_0$) of the quantum well wave function in the z-direction which is smaller than the physical well width and the magnetic length $l_B$ for B ≤ $B_0$ = 16.5 T. The length $l_B$ is closer to $z_0$ for B stronger than $B_0$. The regime $l_B \sim a_0$ (but $l_B > z_0$ ), as could be inferred from above, roughly corresponds to integer quantum Hall regime(IQHR), whereas $l_B < a_0$ (and $l_B$ closer to $z_0$) corresponds to the fractional Hall regime. It follows that, whereas the wave function overlap would be more relevant than the electron-electron repulsion in IQHR, it is just the opposite for FQHR. The numbers introduced and the description given above though correspond to a coarse-grained portrayal of realistic scenario but roughly indicative of the dividing line between the integer and the fractional quantum Hall regimes. In the latter, we compute the viscosity ($\eta$) (i) for different values of the magnetic field strength (B) for given (V,$U_0$), and (ii) for different values of $U_0$ for given ((V,B). A plot of $\eta$ as a function of B in Fig.2 shows that the electronic (dissipative) viscosity, like the dissipation-less Hall viscosity ($\eta_H$)[4,8], increases with B in FQHR. A plot of $\eta$ as a function of $U_0^{-1}$ is shown in Fig.3 for the filling factor $\nu$ =1/3. The plot shows that the dissipation-less state($\eta \to 0^+$), somewhat analogous to super-fluidity, is possible for a value of $U_0$ = 16.66 meV. Since the problem corresponds to a complicated many-body regime, something meaningful about this curious hydrodynamic behavior could be reported only on the basis of further careful investigation, particularly, regarding the nature of collective mode. It may be mentioned in this context that Kinast et al.[ J Kinast et al., Phys. Rev. Lett. **92** 150402(2004)] have shown collective oscillations by confining a gas of lithium-6 atoms in a magneto-optical trap reducing the temperature of the gas by using evaporative cooling technique. The main motivation of the paper is to report the findings shown graphically in Figs.1 and 2. The other one is to show that the KSS bound[3,4] $\eta/S \geq \hbar/4\pi k_B$ derived for all relativistic quantum field theories at finite temperature and zero chemical potential is also valid for the present non-relativistic quantum field theory at finite temperature with non-zero chemical potential.

The paper is organized as follows: In section 2 we focus on the (momentum)spatio-(imaginary)temporal formulation of the MQW problem in the presence of the electron-electron repulsion and magnetic field. We calculate the chemical potential of the fermion number for the filling fraction $\nu = 1/3, 3/7,...$. This is followed by the calculation of entropy and viscosity. The communication ends with concluding remarks in section 3.

## 2 Entropy and viscosity

In this section we wish to present a (momentum)spatio-(imaginary)temporal formulation of the problem involving the Landau level split sub-band energies, the wave function overlaps, and the electron-electron repulsion leading to Matsubara propagators. We also consider the Dyson's equation to pave the way for the inclusion of the disorder/ impurity potential. With these inputs we calculate the chemical potential of the fermion number for different values of the magnetic field strength followed by the calculation of entropy and viscosity.

**2.1 Momentum space Hamiltonian and calculation of Chemical potential**

We start with the second-quantized Hamiltonian $H_{MQW}$ mentioned in section 1, where

$$H_{MQW} = \sum_{k,\alpha,N,\sigma} E_{\alpha,N}(k,B)\, a^\dagger_{k,\alpha,N,\sigma}\, a_{k,\alpha,N,\sigma} + (Ad)^{-1}\sum_{k,k',q,\alpha,N} U_N\, a^\dagger_{k+q,\alpha,N,\uparrow}\, a_{k,\alpha,N,\uparrow}$$

$$a^\dagger_{k'-q,\alpha,N,\downarrow}\, a_{k',\alpha,N,\downarrow}, \qquad (1)$$

$$E_{\alpha,N}(k,B) = [\varepsilon_\alpha + \hbar\omega_c (N+1/2) - 2t_{\alpha,N}(\cos(k_x a) + \cos(k_y a)) + 4 t'_{\alpha,N} \cos(k_x a)\cos(k_y a)]. \qquad (2)$$

In the energy dispersions $E_{\alpha,N}$ above, the Landau levels with each sub-band $\varepsilon_\alpha$ ($\alpha = 0,1$) are indexed by $N=0,1$; **k** is the Bloch wave vector in the first Brillouin zone (BZ). The third and the fourth terms in (2) have their origin in the intra sub-band electron wave functions overlap; the wave functions carry the same Landau level index. The terms corresponding to the intra sub-band and the inter sub-band wave function overlaps, where the electron wave functions carry indices ($\alpha$,N) in such a way that either $\alpha$, or N, or both will be different and electrons are belonging to a particular cell and its neighbouring ones, have not been considered here for. Furthermore, there is no overlap between even and odd parity states for the same well-layer index (j) of the MQW structure under consideration; there is, of course, overlap between even-even and odd-odd parity states for the same well-layer index. However, if one still wishes to consider the inter sub-band overlap between even-odd states, say, between α = 0 and β =1, one should have z≠z´. All such terms have not been considered here. We have taken into consideration only intra-cell electron-electron repulsion ($U_0$, $U_1$). The analysis of the wave function overlap terms, alluded to in section 1, gives $U_1/U_0 \approx 0.7$. In this sub-section we wish to calculate the chemical potential within the Hubbard approximation[5] framework.

The first step of the approximation scheme involves the calculation of (imaginary) time evolution of the operators $a_{k,\alpha,N,\sigma}(\tau)$ where, in units such that $\hbar = 1$, $a_{k,\alpha,N,\sigma}(\tau) = \exp(H_{MQW}\tau)\, a_{k,\alpha,N,\sigma} \exp(-H_{MQW}\tau)$. For the operator $a_{k,0,0,\sigma}(\tau)$ we obtain

$$(\partial/\partial\tau)\, a_{k,0,0,\sigma}(\tau) = -E_{0,0}(k,B)\, a_{k,0,0,\sigma}(\tau) - U_0\, a_{k,0,0,\sigma}(\tau) \sum_{k'} n_{k',0,0,-\sigma}(\tau). \quad (3)$$

At this point we introduce two thermal averages determined by $H_0$, viz. $G_{0,0,\sigma}(k,\tau) = -\langle T\{a_{k,0,0,\sigma}(\tau)\, a^\dagger_{k,0,0,\sigma}(0)\}\rangle$ and $\Gamma_{0,0,\sigma}(k,\tau) = -\sum_{k'} \langle T\{a_{k,0,0,\sigma}(\tau)\, n_{k',0,0,-\sigma}(\tau)\, a^\dagger_{k,0,0,\sigma}(0)\}\rangle$. As the next step, upon using (3), we find that the equations of motion (EOM) of these averages are given by

$$(\partial/\partial\tau)\, G_{0,0,\sigma}(k,\tau) = -E_{0,0}(k,B) G_{0,0,\sigma}(k,\tau) - U_0 \Gamma_{0,0;\sigma}(k,\tau) - \delta(\tau),$$

$$(\partial/\partial\tau)\Gamma_{0,0;\sigma}(k,\tau) \approx -(E_{0,0}(k,B) + U_0 \sum_{k'}\langle n_{k',0,0,-\sigma}\rangle)\times\Gamma_{0,0,\sigma}(k,\tau) - E_{0,0}(k,B)G_{0,0,\sigma}(k,\tau)$$

$$-\sum_{k'}\langle n_{k',0,0,-\sigma}\rangle\, \delta(\tau) \quad (4)$$

where $n_{k,0,0,-\sigma} = a^\dagger_{k,0,0,-\sigma} a_{k,0,0,-\sigma}$. The final step is the calculation of the Fourier coefficients of these temperature Green's functions. We find that

$$G_{0,0,\sigma}(k,\omega_n) = a_{0,0}^{(+)}(k,B)\, (i\omega_n - \acute{\varepsilon}^{(+)}_{0,0}(k,B))^{-1} + a_{0,0}^{(-)}(k,B)\, (i\omega_n - \acute{\varepsilon}^{(-)}_{0,0}(k,B))^{-1} \quad (5)$$

where the coherence factors are given by $a_{0,0}^{(\pm)}(k,B) = (1/2)[1\pm\xi_{0,0}(k,B)]$. The quantity $\xi_{0,0}(k,B)$ and $\acute{\varepsilon}^{(\pm)}_{0,0}(k,B)$ are given by

$$\xi_{0,0}(k,B) = [1 + (4|E_{0,0}(k,B)|/(U_0(\sum_{k'}\langle n_{k',0,0,-\sigma}\rangle)^2))]^{-1/2},$$

$$\acute{\varepsilon}^{(\pm)}_{0,0}(k,B) = E_{0,0}(k,B) + (U_0/2)(\sum_{k''}\langle n_{k'',0,0,-\sigma}\rangle) \times (1 \pm \xi_{0,0}(k,B)^{-1}). \quad (6)$$

In this high energy single-particle spectrum of the system, where the electron-electron interactions set the energy scale as advocated by Jain[9] we find from (19) that the coulomb interactions break the degeneracy of the Landau levels[10] extended possibly across a wide range of Landau level filling fractions. In addition, we notice that the energy gap $\Delta_{k,B} = \acute{\varepsilon}^{(+)}_{0,0}(k,B) - \acute{\varepsilon}^{(-)}_{0,0}(k,B) = U_0(\sum_{k''}\langle n_{k'',0,0,-\sigma}\rangle)\times\xi_{0,0}(k,B)^{-1}$. At low carrier densities, the system may therefore become an insulator with a magnetic-field-tunable energy gap. The tunability aspect has its origin in the term $|E_{0,0}(k,B)|$ which involves magnetic field. In the similar manner we calculate the rest of the propagators $G_{0,1,\sigma}(k,\omega_n)$, $G_{1,0,\sigma}(k,\omega_n)$, and $G_{1,1,\sigma}(k,\omega_n)$. We shall now confine ourselves to the case of the absence of magnetization, i.e. the relative excess of electrons of one spin type being zero. We shall outline how we have included the effect of the elastic scattering by impurities in the propagators $G_{\alpha,N}(k,\omega_n)$.

In view of ref.[6] (see also Appendix (A)), the first order contribution to self-energy is $\Sigma^{(1)}(k,\omega_n) = -(i/2\tau_k) + \Sigma_e$, where $(1/\tau_k) = 2\pi N_j \rho_0 \sum_{k'} |V(k-k')|^2$, $N_j$ is the impurity concentration, and $V(k-k')$ characterizes the momentum dependent impurity potential. The quantity $\tau_k$, in reciprocal energy unit, corresponds to the momentum-dependent life-time of a quasiparticle. The full Matsubara propagators are now given by the Dyson's equation: $G^{(Full)}_{\alpha,N}(k,\omega_n) \approx G_{\alpha,N}(k,\omega_n)/[1-G_{\alpha,N}(k,\omega_n)\Sigma^{(1)}(k,\omega_n)]$. After some straight-forward though tedious algebra (see Appendix (A)), we find that

$$G^{(Full)}_{\alpha,N}(k,\omega_n) = \sum_{j=\pm} a_{ren,k,\alpha,N}^{(j)} (i\omega_n - \acute{\varepsilon}_r^{(j)}{}_{\alpha,N}(k) + i(1/4\tau_{k,\alpha,N}^{(j)}))^{-1} \quad (7)$$

where $a_{ren,k,\alpha,N}^{(j)}, \acute{\varepsilon}_r^{(j)}{}_{\alpha,N}(k)$ and $(1/\tau_{k,\alpha,N}^{(j)})$ are given by (A.8) and (A.9). The un-renormalized single-particle excitation spectrum and the Bogoluibov coherence factors appearing in these terms are given by

$$\acute{\varepsilon}_{\alpha,N}^{(\pm)}(k) = [-|E_{\alpha,N}| + (1/4)U_N n_{\alpha,N} \times \{1 \pm \sqrt{(1+16|E_{\alpha,N}|/U_N n_{\alpha,N}^2)}\}],$$

$$a_{k,\alpha,N}^{(\pm)} = (1/2)[1 \pm (1+16|E_{\alpha,N}|/U_N n_{\alpha,N}^2)^{-1/2}],$$

$$n_{\alpha,N} = \sum_{k,\sigma} \langle a^\dagger_{k,\alpha,N,\sigma} a_{k,\alpha,N,\sigma} \rangle. \quad (8)$$

As shown in Appendix (B) (see Eq. (B.5)), the dimension-less Fermi energy density of states (DOS), involving the disorder broadened Landau level split $\varepsilon_\alpha$'s, can be written in the form

$$\rho_{\alpha,N}^{(\pm)}(\mathbf{k},\omega=\mu) = (1/2\pi^2\rho_0)(\text{Re } a_{ren,k\,\alpha,N}^{(\pm)} \gamma^{(\pm)}_{\mathbf{k},\alpha,N}) \times [(\mu - \acute{\varepsilon}_r^{(\pm)}{}_{\alpha,N}(\mathbf{k}))^2 + \gamma^{(\pm)}_{\mathbf{k},\alpha,N}{}^2]^{-1} \quad (9)$$

where $\gamma_{\mathbf{k},\alpha,N}^{(\pm)} \sim \tau_{\mathbf{k},\alpha,N}^{(\pm)-1}/4$ (the level-broadening factors). At this stage, assuming low concentration of impurities, one may include the contributions of all such diagrams in ref.[6] which involve only one impurity vertex. This gives the equation to determine the total self-energy $\Sigma_{\alpha,N}(\mathbf{k},\omega_n)$:

$$\Sigma_{\alpha,N}(k,\omega_n) = N_j \sum_q V(q) G_{\alpha,N}(k-q,\omega_n) \Gamma_{\alpha,N}(k,q,\omega_n) \quad (10)$$

where $\Gamma_{\alpha,N}(\mathbf{k},\mathbf{q},\omega_n)$ is given by the Lippmann-Schwinger integral equation

$$\Gamma_{\alpha,N}(k,q,\omega_n) = V(-q) + \sum_{q'} V(q'-q) G_{\alpha,N}(k-q',\omega_n) \times \Gamma_{\alpha,N}(k,q',\omega_n). \quad (11)$$

This corresponds to the t-martix approximation. Upon using the optical theorem for the t-matrix[6] one may write

$$\Sigma_{\alpha,N}(\mathbf{k},\omega_n) = i \text{ Im } \Gamma_{\alpha,N}(k,k,\omega_n) = -i\omega_n/(2|\omega_n|\acute{\Gamma}_{k,\alpha,N}) \quad (12)$$

where $\acute{\Gamma}_{k,\alpha,N}^{-1} = 2\pi N_j \rho_0 \sum_{k'} |\Gamma_{\alpha,N}(\mathbf{k},\mathbf{k}')|^2$. Thus the effect of the inclusion of contibution of all the above mentioned diagrams is to replace the Born approximation for scattering by the exact scattering cross-section for a single impurity, i.e. $\tau_k^{-1} \rightarrow \acute{\Gamma}_k^{-1}$. Since $G_{\alpha,N}(\mathbf{k},\omega_n)$ and V(**q**) are specified, using Eqs. (9), (10) and (11) one can determine $\acute{\Gamma}_{k,\alpha,N}^{-1}$ in terms of V(**k**). We note that the procedure outlined has been followed in the numerical calculations in Appendix (C) for the disorder potential $|v_0|$ equal to 4 meV. In the first approximation, we find $\acute{\Gamma}_{k,\alpha,N}^{-1} = 3.8979$ meV which is independent of (k,α,N).

From the Fourier coefficients above, we obtain the thermal averages determined by $H_{MQW}$ in (8), viz. $\langle a^{\dagger}_{k,\alpha,N,\sigma} a_{k,\alpha,N,\sigma}\rangle$, with the inclusion of the chemical potential μ. These averages involve renormalized single-particle excitation spectrum, and the DOSs at ω = μ. With the aid of these averages we may write

$$n_{\alpha,N} = \sum_{k,\sigma} \langle a^{\dagger}_{k,\alpha,N,\sigma} a_{k,\alpha,N,\sigma}\rangle = \sum_{j=\pm} \int d(ka)\, \rho_{\alpha,N}^{(\pm)}(k,\omega=\mu)\times(\exp\beta(\acute{\varepsilon}_r^{(j)}{}_{\alpha,N}(k,B)-\mu)+1)^{-1}, \quad (13)$$

where $\int d(ka) \to \int_{-\pi}^{+\pi}(d(k_x a)/2\pi) \times \int_{-\pi}^{+\pi}(d(k_y a)/2\pi)$. These equations together with the equation which is the analog of the Luttinger sum-rule, for the situation when the magnetic field is kept fixed and the carrier density is changed, viz.

$$2(N_A A) = n_{0,0} + n_{0,1} + n_{1,0} + n_{1,1} \quad (14)$$

constitute the set of self-consistent equations to determine ($n_{0,0}$, $n_{0,1}$, $n_{1,0}$, $n_{1,1}$, μ). We have found that ($N_A A$) = 10. As it is clear from above that the element of self-consistency arises from the fact that the single-particle excitation spectra $\acute{\varepsilon}_r^{(\pm)}{}_{\alpha,N}(k)$ determine as well as determined by ($n_{0,0}$, $n_{0,1}$, $n_{1,0}$, $n_{1,1}$, μ). As already stated in section 1, the maximum number of particles with spin s = ½ per Landau level(LL) is given by D = (20/ν) with the number of current carrying states equal to 10j for the entire MQW system with j well-layers. We wish to consider the cases where ν is an improper fraction such as ν = 1/3, 3/7,.... All the available current carrying states are then stationed at the lowest Landau level for j = 2 well-layers. This renders the job of determining the chemical potential μ a simple exercise as one may take $n_{0,0}$ = 20 and $n_{0,1} = n_{1,0} = n_{1,1} = 0$. In view of (14) we find that the equation to determine μ is given by 2($N_A A$) = $n_{0,0}$. However, when ν = 2, the equation is 2($N_A A$) = $n_{0,0} + n_{0,1}$ where $n_{0,0} = n_{0,1}$ = 10 and $n_{1,0} = n_{1,1} = 0$. We have determined μ for the moderately strong disorder potential ($|v_0| \approx 4.0$ meV) and $U_0$ = 5 meV at T = 3 K when the filling fraction (ν) assume the principal series values 1/3, 3/7, etc.. These values of μ are displayed in table 1 below. For example, we find μ = − 260.39 meV and − 261.04 meV for ν = 1/3 and 3/7, respectively. These values are above the renormalized energy $\acute{\varepsilon}_r^{(-)}{}_{0,0}$ (∼ − 290 meV). In table 2 we have displayed values of μ calculated for $U_0$ = 10 meV. Having determined the chemical potential, we now wish to calculate the entropy per particle and viscosity.

## 2.2 Electronic viscosity

Following the Kadanoff-Baym approach[11], the thermodynamic potential may be given by the expression

$$\Omega(T,B,\mu) = \Omega_0(B) - 2\beta^{-1}\sum_{j,k,\alpha,N}\{\ln\cosh(\beta(\acute{\varepsilon}_r^{(j)}{}_{\alpha,N}(k)-\mu)/2)\} \quad (15)$$

where $\Omega_0(B) = \sum_{j,k,\alpha,N} ((\acute{\varepsilon}_r^{(j)}{}_{\alpha,N}(k) - \mu)$ and $\beta = (k_B T)^{-1}$. The entropy per particle is given by $S = (k_B \beta^2/2N_A A) \times (\partial\Omega/\partial\beta)$ which we can write as $S = (k_B/N_A A) \times s_{dimensionless}$ where $s_{dimensionless} = \sum_k s(k,B)$ and

$$s(k,B) = \sum_{j,\alpha,N} [\ln(\exp(-\beta(\acute{\varepsilon}_r^{(j)}{}_{\alpha,N}(k)-\mu)) + 1) + (\beta(\acute{\varepsilon}_r^{(j)}{}_{\alpha,N}(k)-\mu) + \beta^2(\partial(\acute{\varepsilon}_r^{(j)}{}_{\alpha,N}(k)-\mu)/\partial\beta))$$
$$\times (\exp(\beta(\acute{\varepsilon}_r^{(j)}{}_{\alpha,N}(k)-\mu))+1)^{-1}]. \qquad (16)$$

For the present non-relativistic quantum field theory at finite temperature with non-zero chemical potential, applying the Heisenberg uncertainty principle, we shall now see what does the KSS bound $\eta/S \geq \hbar/4\pi k_B$ (= $6.08 \times 10^{-13}$ K-s) imply for the moderately strong disorder regime. The electronic viscosity ($\eta$) of our MQW system, in the same units as those of Kovtun, Kimura, and others[3,4] may be defined as $\eta = \hbar Q$ where

$$Q = \int d(ka) \sum_{j,\alpha,N} (\acute{\varepsilon}_r^{(j)}{}_{\alpha,N}(k) - \mu) \times \rho_{\alpha,N}^{(j)}(k,\omega=\mu) \tau_{k,\alpha,N}^{(j)}, \qquad (17)$$

Upon ignoring the momentum dependence of $\tau_{k,\alpha,N}^{(j)}$ in the first approximation, we can write $\eta \approx \hbar \ddot{\varepsilon} \tau$, where $\ddot{\varepsilon} = \int \sum_{j,\alpha,N} (\acute{\varepsilon}_r^{(j)}{}_{\alpha,N}(\mathbf{k})-\mu) \rho_{\alpha,N}^{(j)}(\mathbf{k},\omega=\mu) d(\mathbf{k}a)$. Note that, defined in this manner, $\ddot{\varepsilon}$ (and therefore $\eta$) can assume positive as well as negative values depending upon the location of the chemical pontential within the two sub-bands under consideration. Now according to the uncertainty principle, for $\ddot{\varepsilon} > 0$, the product of the energy of a quasiparticle ($\ddot{\varepsilon}/2N_A A$) and ($\tau \hbar$) cannot be smaller than $\hbar$, otherwise the quasiparticle concept does not make sense. Therefore we obtain, from the uncertainty principle alone, that $\eta \geq 2(N_A A)\hbar$. Similarly, for $\ddot{\varepsilon} < 0$, $\eta \leq -2(N_A A)\hbar$. These inequalities are satisfied easily. The entropy per particle (S), on the other hand, is ($k_B/N_A A$) $s_{dimensionless}$. Therefore, for $\ddot{\varepsilon} > 0$, $\eta/S \geq 2k_B^{-1}(N_A A)^2 \hbar/s_{dimensionless}$ which implies that the quantity $\eta/S \geq \hbar/4\pi k_B$ provided the dimensionless quantity

$$(N_A A)^{-2} s_{dimensionless} \leq 8\pi. \qquad (18)$$

Likewise, for $\ddot{\varepsilon} < 0$, $\eta/S \leq -2k_B^{-1}(N_A A)^2 \hbar/s_{dimensionless}$ which implies that the quantity $\eta/s \leq -\hbar/4\pi k_B$ provided the quantity

$$(N_A A)^{-2} s_{dimensionless} \geq 8\pi. \qquad (19)$$

In our quantum confined system, as we infer from the values of $s_{dimensionless}$ given in tables 1 and 2, that the inequality (18) is easily satisfied for the moderately strong disorder potential, for the number of current carrying states in the single well-layer area A is ($N_A A$) = 10. This indicates the possible universality of the inequality $\eta/s \geq \hbar/4\pi k_B$.

**Table1** The summary of the values of $\mu$, $s_{dimensionless}$, $\eta$, and the thermodynamic potential ($\Omega(T,B,\mu)$) obtained in the presence of moderate disorder(4 meV) and temperature $T = 30$ K for $U_0 = 5$ meV. The entropy and $\Omega$ are found to increase with decrease in $\nu$.

| Filling Factor (ν) | The chemical potential (μ) in meV | The dimension-less Entropy($s_{dimensionless}$) | The dimension-less electronic viscosity(Q) | Ω(T,B,μ) in meV |
|---|---|---|---|---|
| 1/3 | −260.39 | 2.1435 | 755.66 | −3.0150×10$^6$ |
| 3/7 | −261.04 | 2.1280 | 753.69 | −3.0170×10$^6$ |
| ½ | −261.36 | 2.1166 | 752.93 | −3.0185×10$^6$ |
| 2/3 | −261.86 | 2.1145 | 750.89 | −3.0188×10$^6$ |
| 1 | −262.35 | 2.1050 | 749.31 | −3.0200×10$^6$ |

**Table 2** The summary of the values of μ, $s_{dimensionless}$, η , and Ω(T,B,μ) obtained in the presence of moderately strong disorder (4 meV) and temperature T = 30 K for $U_0$ = 10 meV. The Kovtun-Kimura inequality is not violated here.

| Filling Factor (ν) | The chemical potential (μ) in meV | The dimension-less Entropy ($s_{dimensionless}$) | The dimension-less electronic viscosity(Q) | Ω(T,B,μ) in meV |
|---|---|---|---|---|
| 1/3 | −268.49 | 14.5294 | 178.1791 | −2.4758×10$^6$ |
| 3/7 | −269.14 | 14.3840 | 177.4380 | −2.4786×10$^6$ |
| ½ | −269.47 | 14.3371 | 176.9370 | −2.4795×10$^6$ |
| 2/3 | −269.97 | 14.2922 | 176.0552 | −2.4804×10$^6$ |
| 1 | −270.46 | 14.1974 | 175.4376 | −2.4823×10$^6$ |

As mentioned earlier, our approach involves calculation of the density of viscosity η(k), for temperature close to 0 K, on the Brillouin zone(BZ) (see Fig.1) followed by evaluation of the integral of η(k) over the BZ. Upon evaluating the integral, we obtain the viscosity η. We have shown a plot of (η/ℏ) as a function of the magnetic field strength (eB/ $N_A$ h) in Fig.2 using the values displayed in table1. The plot almost corresponds to a straight line. The result is not counter-intuitive, as a strong magnetic field will encourage localization and naturally more resistance to flow. Furthermore, a plot of (η/ℏ) as a function of $U_0^{-1}$ for filling factor (1/3) in Fig.3 shows that the dissipation-less state(η→0$^+$) (inviscid fluid), somewhat analogous to super-fluidity, is possible for a value of $U_0$ = 16.66 meV. We have not observed any discontinuity in the entropy or the specific heat (C = −β (∂S/∂β)) given by

$$C \propto k_B \sum_{k,j,\alpha,N} [\beta (\acute{\varepsilon}_r^{(j)}{}_{\alpha,N}(k) - \mu)]^2 \times \exp(\beta (\acute{\varepsilon}_r^{(j)}{}_{\alpha,N}(k) - \mu))$$

$$\times (\exp(\beta(\acute{\varepsilon}_r^{(j)}{}_{\alpha,N}(k) - \mu)) + 1)^{-2}. \qquad (20)$$

At this stage, therefore, η→0$^+$ may be termed as a cross-over (and not a phase transition).

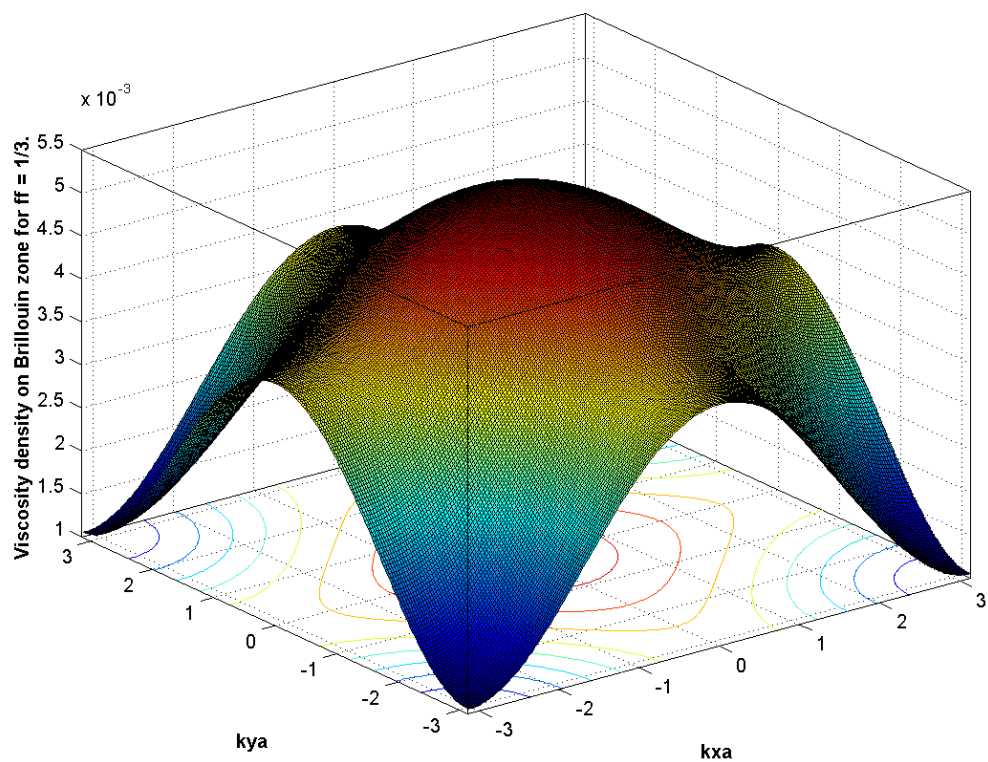

**Figure 1** A 3-D plot of the electronic viscosity density on the first Brillouin zone for the filling factor (ff) = 1/3 and the electron-electron repulsion $U_0 = 10$ meV.

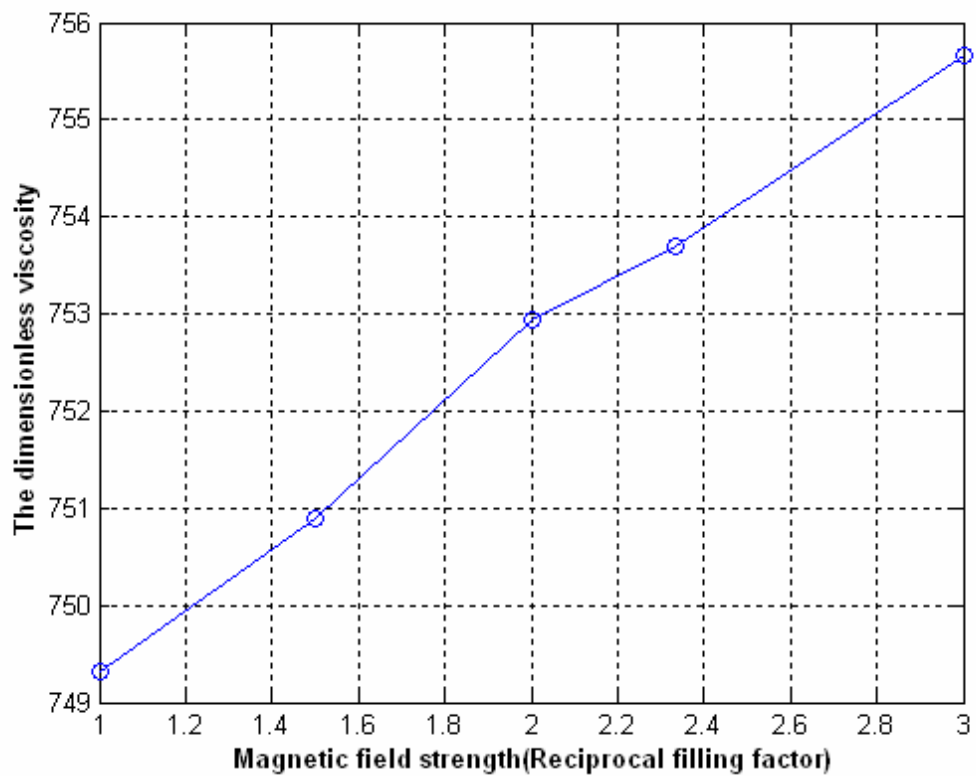

**Figure2** A 2-D plot of the dimensionless viscosity as a function of the magnetic field strength(reciprocal filling factor $\nu^{-1}$)for the electron-electron repulsion $U_0$ = 5 meV, and the disorder potential $|v_0| \approx 4$ meV. The plot is very nearly a straight line.

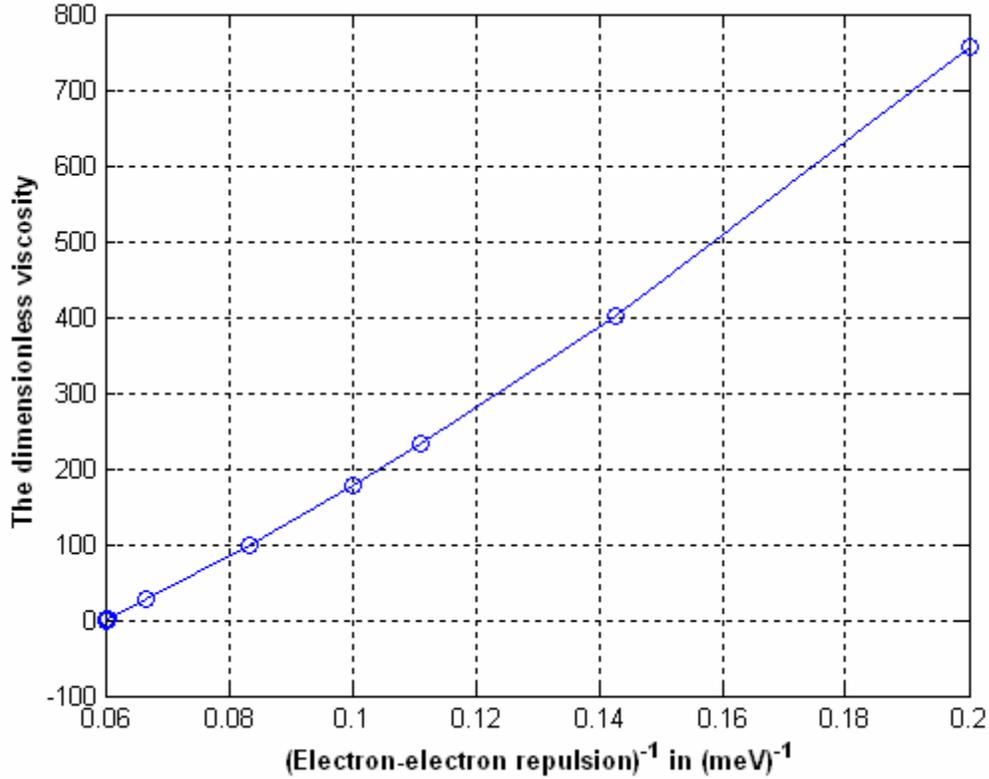

**Figure3** A 2-D plot of the dimensionless viscosity as a function of the (Electron-electron repulsion)$^{-1}$ for the filling fraction $\nu$ =1/3, and the disorder potential $|v_0| \approx 4$ meV. The plot shows that the dissipation-less state($\eta \rightarrow 0^+$), somewhat analogous to super-fluidity, is possible for a value of $U_0$ = 16.66 meV.

The reader may note that we have not discarded the possibility of the counter-intuitive result, viz. negative η where (19) is satisfied. The negative η obtained here for ν = 2 (μ = − 286.03 meV, $s_{dimensionless}$ = 149.10, Q = −148.97) does not comply with (19) and therefore to be discarded. As pointed out in section 1 and sub-section 2.1, for the investigation in the integer quantum Hall regime, we need to include the intra-/inter sub-band wave function overlaps where the electron wave functions carry indices (α,N) in such a way that either α, or N, or both will be different. Since this inclusion has not been done here, the result for ν = 2 may not carry much meaning. It may be mentioned in passing that negative η is, though unusual, not unheard of. For the Newtonian magnetic fluids, in an alternating linearly polarized magnetic field, the negative viscosity effect has already been observed and reported by Bacri et al.[12] several years ago. Besides, it may also be noted that η < 0 bears a clear analogy with the negative magneto-resistance effect[13] known to occur in the "weakly" localized system.

3 **Concluding remarks**  We have derived and studied a Green's function matrix involving for a multiple quantum well structure including electron-electron interaction in the presence of disorder potential and a magnetic field (B) in the direction in which the electrons are trapped at a finite temperature. The matrix led to the derivation of expressions for viscosity. Within the frame-work of this formulation, electronic specific heat, conductivity, etc. for the structure could also be calculated. Our investigation shows that the electron-electron interaction $U_0$ is a crucial parameter. A change in $U_0$ leads to a different set of numerical values for the viscosity as shown in tables 1 and 2 with η nearly proportional to B. The formulation and findings in this paper are provocative but far from comprehensive, as a suitable scheme is to be devised to tackle sub-band hoppings, Landau level mixing and intra-/inter-cell electron-electron repulsion, all together in a meaningful manner, for the fractional as well as integer quantum Hall regimes and, though leaning heavily on numerics, the finding η nearly proportional to B seems to be fairly general for MQW systems needing experimental confirmation. In conclusion, a system-specific theoretical investigation, such as ours, eventually has to find points of convergence with the seminal works [14,15,16,17] on quantum Hall fluids, involving explicit wave function construction, for its wider acceptability. This is an important task ahead of us.

**Acknowledgements** It is a pleasure to thank Taro Kimura for sparing time to read the manuscript of this paper in preparatory stage and making us aware, among others, a crucial fact that, no matter how one calculates the viscosity coefficient, the KSS bound should not be violated. This helped us to discard the negative viscosity ( not complying with the KSS bound) we obtained in certain cases.

## Appendix (A)

Consider the propagator

$$G_{\alpha,N}(k, \omega_n) = a_{k,\alpha,N}^{(+)}(k,B)\,(i\omega_n - \acute{\varepsilon}^{(+)}_{\alpha,N}(k, B))^{-1} + a_{k,\alpha,N}^{(-)}(\mathbf{k},B)\,(i\omega_n - \acute{\varepsilon}^{(-)}_{\alpha,N}(\mathbf{k}, B))^{-1}$$

(A.1)

Assuming the scattering by impurities weak, as in ref.[6], we may write $\Sigma^{(1)}(\mathbf{k},\omega_n) = N_j \sum_{\mathbf{k}'} |V(\mathbf{k}-\mathbf{k}')|^2 G_{\alpha,N}(\mathbf{k}',\omega_n) = \Sigma_0^{(1)}(\mathbf{k},\omega_n) + \Sigma_e$ where

$$\Sigma_0^{(1)}(\mathbf{k},\omega_n) = - N_j \sum_{\mathbf{k}'} |V(\mathbf{k}-\mathbf{k}')|^2 (i\omega_n) \int_{-\infty}^{+\infty} d\varepsilon\, \rho(\varepsilon)$$

$$\times [a_{k,\alpha,N}^{(+)} (\omega_n^2 + \acute{\varepsilon}^{(+)2}_{\alpha,N})^{-1} + a_{k,\alpha,N}^{(-)} (\omega_n^2 + \acute{\varepsilon}^{(-)2}_{\alpha,N})^{-1}] \quad (A.2)$$

and $\Sigma_e$ is the part of the first order contribution which can be shown to be independent of $\mathbf{k}$ and $\omega_n$ for $\mathbf{k}$ close to Fermi momentum and can be absorbed in chemical potential. To evaluate the integrals in (A.2), such as $\int_{-\infty}^{+\infty} d\varepsilon\, \rho(\varepsilon) (\omega_n^2 + \acute{\varepsilon}^{(+)2}_{\alpha,N})^{-1}$, we assume $\rho(\varepsilon) = \rho_0 \delta(\varepsilon - \acute{\varepsilon}^{(+)}_{\alpha,N})$ where $\rho_0 \sim 0.1$ (meV)$^{-1}$. This yields

$$G_{\alpha,N}(\mathbf{k},\omega_n) \approx - \rho_0 (i\omega_n)(\pi/|\omega_n|), \quad \Sigma_0^{(1)}(\mathbf{k},\omega_n) = - N_j \rho_0 (i\omega_n) \sum_{\mathbf{k}'} |V(\mathbf{k}-\mathbf{k}')|^2 (\pi/|\omega_n|). \quad (A.3)$$

We may write the right-hand side of the second equation in (A.3) as $[-i\omega_n /(2|\omega_n|\tau_k)]$, where $(1/\tau_k) = 2\pi N_j \rho_0 \sum_{\mathbf{k}'} |V(\mathbf{k}-\mathbf{k}')|^2$. Note that $\tau_k$, which corresponds to quasi-paticle lifetime(QPLT), is expressed in reciprocal energy units. We now wish to show how (7) comes about.

Upon using the Dyson's equation, the full propagator may be written as $G^{(Full)}_{\alpha,N}(k, \omega_n) \approx D/E$, where

$$D = [a_{\alpha,N}^{(+)} (i\omega_n - \acute{\varepsilon}^{(-)}_{\alpha,N}) + a_{\alpha,N}^{(-)} (i\omega_n - \acute{\varepsilon}^{(+)}_{\alpha,N})],$$

$$E = (i\omega_n)^2 - (i\omega_n)(\acute{\varepsilon}^{(+)}_{\alpha,N} + \acute{\varepsilon}^{(-)}_{\alpha,N} - (i/2\tau_k) + \Sigma_e)$$

$$+ \{\acute{\varepsilon}^{(+)}_{\alpha,N} \acute{\varepsilon}^{(-)}_{\alpha,N} + (-(i/2\tau_k) + \Sigma_e)(a_{\alpha,N}^{(+)} \acute{\varepsilon}^{(-)}_{\alpha,N} + a_{\alpha,N}^{(-)} \acute{\varepsilon}^{(+)}_{\alpha,N})\}. \quad (A.4)$$

We have dropped the argument part from the coherence factors and the single-particle excitation spectra above for convenience. The roots of the equation $E = 0$ are

$$i\omega_n = \{(\acute{\varepsilon}^{(+)}_{\alpha,N} + \acute{\varepsilon}^{(-)}_{\alpha,N} - (i/2\tau_k) + \Sigma_e)/2\} \pm (R_{k,\alpha,N}^{1/2}/2)(\cos(\theta_{k,\alpha,N}/2) - i\sin(\theta_{k,\alpha,N}/2)),$$

$$R_{k,\alpha,N} = [E_1^2 + E_2^2]^{1/2}, \quad \tan(\theta_{k,\alpha,N}) = E_2/E_1,$$

$$E_1 = (\acute{\varepsilon}^{(+)}_{\alpha,N} - \acute{\varepsilon}^{(-)}_{\alpha,N})^2 + \Sigma_e^2 - (1/4\tau_k^2) + 2(\acute{\varepsilon}^{(+)}_{\alpha,N} + \acute{\varepsilon}^{(-)}_{\alpha,N})\Sigma_e - 4\Sigma_e(a_{\alpha,N}^{(+)} \acute{\varepsilon}^{(-)}_{\alpha,N} + a_{\alpha,N}^{(-)} \acute{\varepsilon}^{(+)}_{\alpha,N}),$$

$$E_2 = \{(\acute{\varepsilon}^{(+)}_{\alpha,N} + \acute{\varepsilon}^{(-)}_{\alpha,N} + \Sigma_e)/\tau_k\} - (2/\tau_k)(a_{\alpha,N}^{(+)} \acute{\varepsilon}^{(-)}_{\alpha,N} + a_{\alpha,N}^{(-)} \acute{\varepsilon}^{(+)}_{\alpha,N}). \quad (A.5)$$

It follows that the denominator $E$ of $G^{(Full)}_{\alpha,N}(\mathbf{k}, \omega_n)$ may be written as the product of two factors $(i\omega_n - (\alpha^{(+)} + i\beta^{(+)}))(i\omega_n - (\alpha^{(-)} + i\beta^{(-)}))$ where

$$\alpha^{(\pm)} = \{(\acute{\varepsilon}^{(+)}_{\alpha,N} + \acute{\varepsilon}^{(-)}_{\alpha,N} + \Sigma_e)/2\} \pm (R_{k,\alpha,N}^{1/2}/2)\cos(\theta_{k,\alpha,N}/2),$$

$$\beta^{(\pm)} = -\{(1/4\tau_k) \pm (R_{k,\alpha,N}^{1/2}/2)\sin(\theta_{k,\alpha,N}/2)\}. \quad (A.6)$$

In view of (A.6), $G^{(Full)}(\mathbf{k},\omega_n) \approx D/E$ may be written as

$$G^{(Full)}(\mathbf{k},\omega_n) = a_{ren,k,\alpha,N}^{(+)} [i\omega_n - \acute{\varepsilon}_r^{(+)}{}_{\alpha,N} + i(1/4\tau_{k,\alpha,N}^{(+)})]^{-1}$$
$$+ a_{ren,k,\alpha,N}^{(-)} [i\omega_n - \acute{\varepsilon}_r^{(-)}{}_{\alpha,N} + i(1/4\tau_{k,\alpha,N}^{(-)})]^{-1}. \quad (A.7)$$

where

$$\acute{\varepsilon}_r^{(\pm)}{}_{\alpha,N} = \alpha^{(\pm)}, \quad (1/\tau_{k,\alpha,N}^{(\pm)}) = \{(1/\tau_k) \pm (2R_{k,\alpha,N}^{1/2}) \sin(\theta_{k,\alpha,N}/2)\},$$

$$a_{ren,k,\alpha,N}^{(\pm)} = (1/2)(1 \pm \delta_{k,\alpha,N}), \delta_{k,\alpha,N} = (\delta^{(1)}{}_{k,\alpha,N} / \delta^{(2)}{}_{k,\alpha,N}), \quad (A.8)$$

and

$$\delta^{(1)}{}_{k,\alpha,N} = [(\alpha^{(+)} + i\beta^{(+)})(\alpha^{(-)} + i\beta^{(-)}) - 2(a_{\alpha,N}^{(+)} \acute{\varepsilon}^{(-)}{}_{\alpha,N} + a_{\alpha,N}^{(-)} \acute{\varepsilon}^{(+)}{}_{\alpha,N})],$$

$$\delta^{(2)}{}_{k,\alpha,N} = [(\alpha^{(+)} + i\beta^{(+)}) - (\alpha^{(-)} + i\beta^{(-)})]. \quad (A.9)$$

The renormalized Bogoluibov coherence factors $a_{ren,k,\alpha,N}^{(\pm)}$ turn out to be complex quantities. The implication of this will be clarified in Appendix (B).

**Appendix (B)**

We have calculated explicitly the propagators $G^{(Full)}{}_{\alpha,N}(\mathbf{k}, \omega_n)$ in Appendix (A) with the inclusion of impurity scattering. The corresponding retarded Green's function $G^{(R)}{}_{\alpha,N}(\mathbf{k},t)$, in units such that $\hbar = 1$, is given by $G^{(R)}{}_{\alpha,N}(\mathbf{k},t) = {}_{-\infty}\!\int^{+\infty} (d\omega/2\pi) \exp(-i\omega t) G^{(Full)}{}_{\alpha,N}(\mathbf{k},\omega)$ where in the upper and lower half-plane, respectively,

$$G^{(Full)}{}_{\alpha,N}(\mathbf{k},\omega) = \sum_{j=\pm} a_{ren,k,\alpha,N}^{(j)} (\omega - \acute{\varepsilon}_r^{(j)}{}_{\alpha,N}(\mathbf{k}) + i(1/4\tau_{k,\alpha,N}^{(j)}))^{-1} \quad (B.1)$$

and

$$G^{(Full)}{}_{\alpha,N}(\mathbf{k},\omega) = \sum_{j=\pm} a_{ren,k,\alpha,N}^{(j)} (\omega - \acute{\varepsilon}_r^{(j)}{}_{\alpha,N}(\mathbf{k}) - i(1/4\tau_{k,\alpha,N}^{(j)}))^{-1}. \quad (B.2)$$

Thus $G^{(R)}{}_{\alpha,N}(\mathbf{k},\omega') = {}_{-\infty}\!\int^{+\infty} dt \exp(i\omega' t) G^{(R)}{}_{\alpha,N}(\mathbf{k},t)$ is given by (B.1) with $\omega$ real. We obtain

$$G^{(R)}{}_{\alpha,N}(\mathbf{k},t) = \sum_{j=\pm} a_{ren,k,\alpha,N}^{(j)} i\exp(-i \acute{\varepsilon}_r^{(j)}{}_{\alpha,N}(\mathbf{k})t - (t/4\tau_{k,\alpha,N}^{(j)})) \theta(t) \quad (B.3)$$

where the unit step function $\theta(t) = {}_{-\infty}\!\int^{+\infty} (id\omega/2\pi)\{\exp(-i\omega t) / (\omega + i 0^+)\}$. This shows that the impurity scattering leads to finite lifetime for the fermion states of definite momentum. Using the integral representation of $\theta(t)$ above it is not difficult to show that

$$G^{(R)}{}_{\alpha,N}(\mathbf{k},\omega') = {}_{-\infty}\!\int^{+\infty} dt \exp(i\omega' t) G^{(R)}{}_{\alpha,N}(\mathbf{k},t)$$

$$= \sum_{j=\pm} a_{ren,k,\alpha,N}^{(j)} (\omega - \acute{\varepsilon}_r^{(j)}{}_{\alpha,N}(\mathbf{k}) + i(1/4\tau_{k,\alpha,N}^{(j)})) \times [(\omega - \acute{\varepsilon}_r^{(j)}{}_{\alpha,N}(\mathbf{k}))^2 + (1/4\tau_{k,\alpha,N}^{(j)})^2]^{-1}. \quad (B.4)$$

As the renormalized Bogoluibov coherence factors $a_{ren,k,\alpha,N}^{(\pm)}$ are found to be complex, the dimensionless density of states $\rho_{\alpha,N}(\mathbf{k},\omega) \equiv (-1/2\pi^2\rho_0) \operatorname{Im} G^{(R)}{}_{\alpha,N}(\mathbf{k},\omega)$ comprises of two parts: $\rho_{\alpha,N}(\mathbf{k},\omega) = \rho'_{\alpha,N}(\mathbf{k},\omega) + \rho''_{\alpha,N}(\mathbf{k},\omega)$, where

$$\rho'_{\alpha,N}(\mathbf{k},\omega) = (1/2\pi^2\rho_0) \sum_{j=\pm} (\operatorname{Re} a_{ren,k,\alpha,N}^{(j)}) \gamma^{(j)}_{\mathbf{k},\alpha,N} \times [(\omega - \acute{\varepsilon}_r^{(j)}{}_{\alpha,N}(\mathbf{k}))^2 + \gamma^{(j)}_{\mathbf{k},\alpha,N}{}^2]^{-1}, \quad (B.5)$$

$$\rho''_{\alpha,N}(\mathbf{k},\omega)=(-1/\pi\rho_0)\sum_{j=\pm}(\text{Im } a_{ren,k,\alpha,N}^{(j)})(\omega-\acute{\varepsilon}_r^{(j)}{}_{\alpha,N}(\mathbf{k}))\times[(\omega-\acute{\varepsilon}_r^{(j)}{}_{\alpha,N}(\mathbf{k}))^2+\gamma^{(j)}{}_{\mathbf{k},\alpha,N}{}^2]^{-1}, \quad (B.6)$$

$$\text{Re } a_{ren,k,\alpha,N}^{(\pm)}=(1/2)(1\pm\text{Re } \delta_{k,\alpha,N}), \quad \text{Im } a_{ren,k,\alpha,N}^{(\pm)}=(1/2)(1\pm\text{Im } \delta_{k,\alpha,N}), \quad (B.7)$$

$$\text{Re } \delta_{k,\alpha,N}= \delta'_{k,\alpha,N}/\delta'''_{k,\alpha,N}, \quad \text{Im } \delta_{k,\alpha,N}= \delta''_{k,\alpha,N}/\delta'''_{k,\alpha,N}, \quad (B.8)$$

$$\delta'_{k,\alpha,N} = [(\alpha^{(+)2}- \alpha^{(-)2})+ (\beta^{(+)2}- \beta^{(-)2}) -2(a_{\alpha,N}{}^{(+)}\acute{\varepsilon}^{(-)}{}_{\alpha,N}+ a_{\alpha,N}{}^{(-)}\acute{\varepsilon}^{(+)}{}_{\alpha,N})\times (\alpha^{(+)}- \alpha^{(-)})], \quad (B.9)$$

$$\delta''_{k,\alpha,N} = 2[\alpha^{(+)}\beta^{(-)}-\alpha^{(-)}\beta^{(+)}], \delta'''_{k,\alpha,N} = [(\alpha^{(+)2}- \alpha^{(-)2})+ (\beta^{(+)2}- \beta^{(-)2})], \quad (B.10)$$

and $\gamma_{\mathbf{k},\alpha,N}{}^{(\pm)}= \tau_{\mathbf{k},\alpha,N}{}^{(\pm)-1}/4$ (the level-broadening factors). In order to determine the Fermi energy density of states (DOS) $\rho_{Fermi}(\mathbf{k})$, since we shall put $\omega=\mu$ in (B.5) and (B.6), it is clear that

$$\rho_{Fermi}(\mathbf{k}) = (1/2\pi^2\rho_0) \sum_{j=\pm,\alpha,N} (\text{Re } a_{ren,k,\alpha,N}^{(j)}) \gamma^{(j)}{}_{\mathbf{k},\alpha,N} \times[(\mu- \acute{\varepsilon}_r^{(j)}{}_{\alpha,N}(\mathbf{k}))^2+\gamma^{(j)}{}_{\mathbf{k},\alpha,N}{}^2]^{-1}. \quad (B.11)$$

Equation (B.6) does not contribute here as the branches of the Fermi surface are given by $(\mu- \acute{\varepsilon}_r^{(j)}{}_{\alpha,N}(\mathbf{k})) = 0$. However, for $\omega\neq \mu$, definitely this equation will contribute towards the DOS.

**Appendix (C)**

We now turn our attention to Eqs.(9) and (10). In the limit $\kappa >> |\mathbf{k}-\mathbf{k'}|$, as seen in section 1, the disorder potential $V(|\mathbf{q}|) \approx |v_0|$ and, therefore, from the latter we obtain

$$\Gamma_{\alpha,N}(\mathbf{k},\omega_n) \approx |v_0|/(1- |v_0| G_{\alpha,N}(\mathbf{k},\omega_n)) \quad (C.1)$$

In view of Eq. (A.3), we find that

$$\text{Im } \Gamma_{\alpha,N}(\mathbf{k},\omega_n) \approx - \rho_0 \pi |v_0|^2 /(1+ \rho_0{}^2\pi^2 |v_0|^2 ). \quad (C.2)$$

From Eq.(11) we now find that $\acute{\Gamma}_{k,\alpha,N}{}^{-1}$, in the first approximation, is given by $[2 \rho_0 \pi |v_0|^2 /(1+ \rho_0{}^2\pi^2 |v_0|^2 )]$. We take a moderately strong disorder potential $|v_0| = 4.1380$ meV. This gives $\acute{\Gamma}_{k,\alpha,N}{}^{-1}= 4$ meV. The choice is obviously dictated by the fact that $\gamma_{k,\alpha,N} = \acute{\Gamma}_{k,\alpha,N}{}^{-1}/4$ and, with $\acute{\Gamma}_{k,\alpha,N}{}^{-1}= 4$ meV, the momentum-independent level-broadening factors are equal to 1 meV. The quantity $\Sigma_e$ has been taken to be 0.1 meV for the calculation above. It may be mentioned in passing that one can also consider a weak disorder potential, say, $|v_0| = 0.4$ meV, which gives $\acute{\Gamma}_{k,\alpha,N}{}^{-1}\approx 0.1$ meV – a much longer quasi-particle life time compared to that for $|v_0| = 4$ meV. In this case, however, we obtain negative viscosity– a counter-intuitive result not complying with the KSS bound. We note that, even though $\acute{\Gamma}_{k,\alpha,N}$ is found to be k-independent in the first approximation, the term $\pm 4R_{k,\alpha,N}{}^{1/2} \sin(\theta_{k,\alpha,N}/2)$ in (A.8) will ensure that $\tau_{k,\alpha,N}{}^{(\pm)}$ are momentum dependent and different for the upper and lower branches.